# Single virus fingerprinting by widefield interferometric defocus-enhanced mid-infrared photothermal microscopy

Qing Xia[1], Zhongyue Guo[2], Haonan Zong[1], Scott Seitz[3], Celalettin Yurdakul[1], M. Selim Ünlü[1], Le Wang[1], John H. Connor[3,*] and Ji-Xin Cheng[1,2,4,*]

[1] Department of Electrical and Computer Engineering, Boston University, Boston, Massachusetts 02215, United States;

[2] Department of Biomedical Engineering, Boston University, Boston, Massachusetts 02215, United States;

[3] Department of Microbiology and National Infectious Diseases Laboratories, Boston University School of Medicine, Boston, Massachusetts 02118, United States;

[4] Photonics Center, Boston University, Boston, Massachusetts 02215, United States.

**Abstract**

Clinical identification and fundamental study of viruses rely on the detection of viral proteins or viral nucleic acids. Yet, amplification-based and antigen-based methods are not able to provide precise compositional information of individual virions due to small particle size and low-abundance chemical contents (e.g., ~ 5000 proteins in a vesicular stomatitis virus). Here, we report a widefield interferometric defocus-enhanced mid-infrared photothermal (WIDE-MIP) microscope for high-throughput fingerprinting of single viruses. With the identification of feature absorption peaks, WIDE-MIP reveals the contents of viral proteins and nucleic acids in single DNA vaccinia viruses and RNA vesicular stomatitis viruses. Different nucleic acids signatures of thymine and uracil residue vibrations are obtained to differentiate DNA and RNA viruses. WIDE-MIP imaging further reveals an enriched β sheet components in DNA varicella-zoster virus proteins. Together, these advances open a new avenue for compositional analysis of viral vectors and elucidating protein function in an assembled virion.

**Main**

The emergence of the monkeypox outbreak in early 2022 has posed a new global health threat during the coronavirus 19 (COVID-19) pandemic[1, 2, 3]. With the spread of virus-based infectious diseases, rapid and accurate testing is crucial for mitigating the impact of current and future pandemics[4]. Diagnostic tests on the viruses commonly rely on the detection of nucleic acids or surface proteins. Generally, the amount of viral nucleic acid in a single virion is lower than the amount of viral protein. Detecting viral nucleic acids is challenging without signal amplification techniques such as polymerase chain reaction. Although nucleic acid amplification tests[4, 5, 6] and antigen rapid diagnostic tests[7, 8] can provide accurate testing results, they usually require pre-treatments of a large amount of virions, extraction or tagging that add time to any assay[5]. It is noteworthy that residual viral RNA from patient specimens remains detectable even though patients have recovered or without culturable viruses[9, 10, 11]. Thus, in addition to detecting viral fragments, new complement assays are required to identify the intact virions with preserved structure in order to confirm viral infection and reduce false diagnoses.

Accelerated efforts have been devoted to developing label-free technologies, in which optical detection and morphological characterization of single viruses have shown to be promising for clinical diagnosis[12, 13]. Although the scattering from a single virion is weak, it can be enhanced by interfering with a strong reference field in an interferometric light microscope[14]. With the enhanced signal contrasts, interferometric imaging has been used for single-virus tracking and viral infection study[15, 16, 17]. Towards translation into clinic, interferometric sensing methods have also demonstrated the visualization of single viruses in undiluted fetal bovine serum[18] and rapid detection of single intact virion in human saliva[19]. However, these methods lack molecular information of the viruses while the chemical contents are critical to viral structure and function[20].

Vibrational spectroscopic detection of viruses is valuable for analyzing the chemical components of virus strains[21, 22, 23]. Methods relying on either Raman scattering or infrared (IR) absorption offer intrinsic chemical selectivity at single-virus level by using spectroscopic signatures of chemical bonds[24, 25]. Compared to Raman scattering, IR absorption offers 8 orders of magnitude larger cross section that enables adequate chemical sensitivity and throughput[26]. Mid-infrared photothermal (MIP) microscopy is an emerging technique based on mapping of local transient heat to achieve IR spectroscopic imaging at the diffraction limit of visible light[27, 28]. In MIP microscopy, a visible probe beam is used to detect photothermal-based chemical contrast induced by a mid-IR pump beam[29]. Since the first demonstration of 3D MIP imaging of living cells,[29] MIP microscopy has enabled broad applications in life science, ranging from individual bacteria[30], single cells[31, 32, 33], sliced tissues[34], to entire organisms[35]. With counter-propagation of IR and visible beams, researchers have shown MIP imaging of 100 nm polystyrene beads[36, 37]. With interferometric scattering as the probe in a confocal configuration, MIP spectroscopic detection of a single virus was reported[38]. However, the scanning-based MIP methods suffer from long acquisition time and low throughput. Although widefield MIP imaging was developed to allow ultrafast chemical imaging at a speed up to 1250 frames per second[39], it remains very challenging for widefield MIP to detect single viral nanoparticles and perform precise spectral analysis.

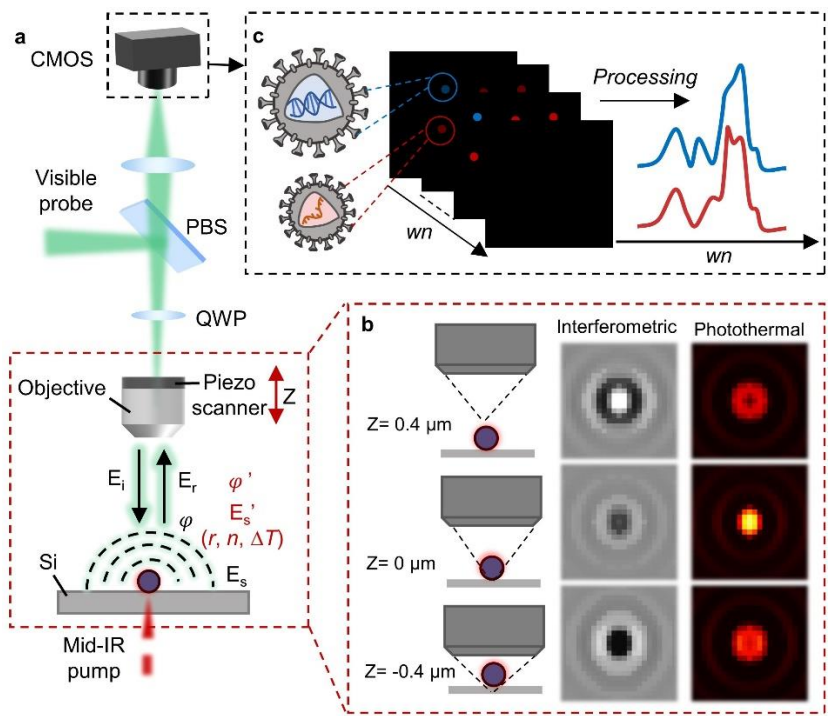

**Fig. 1 Schematic and principle of WIDE-MIP microscopy.** (a) Schematic of a WIDE-MIP microscope and principle of interferometric scattering-based MIP imaging. CMOS: complementary metal-oxide semiconductor. PBS: Polarizing beam splitter, QWP: quarter-wave plate, $E_i$: visible incident light field, $E_r$: reflected field by the substrate, $E_s$: scattered field by the sample, $E_s'$: IR modulation resulted scattered field by the sample, which is related to the radius ($r$), refractive index ($n$) and temperature change ($\Delta T$) of the sample, $\varphi$: phase difference between $E_s$ and $E_r$. A delay pulse generator is used to synchronize the pump pulse, probe pulse and camera (Supplementary Fig. 1). (b) Schematic illustration of interferometric defocus-enhanced photothermal contrast provided by Z-axis scanning of objective. The different positions are with respect to the substrate top surface (Z = 0 μm). (c)

Principle of fingerprinting DNA and RNA viruses by WIDE-MIP. wn: wavenumber. Hyperspectral images of single viruses are recorded by continuously tuning the IR wavenumber. Blue dots indicate DNA viruses, red dots indicate RNA viruses.

Here, we present the development and validation of a widefield interferometric defocus-enhanced MIP (WIDE-MIP) microscope (Fig. 1a) for fingerprint analysis of bio-nanoparticles. As photothermal signal is a modulation of visible beam intensity, it is commonly believed that an optimal MIP contrast is generated when the bright field contrast is maximized. Yet, this wisdom does not hold for interferometric MIP microscopy where the signal strongly depends on the relative phase between the particle-scattered photons and the substrate-reflected reference field. Instead, we find that by fine tuning the focus position of the objective, the defocused interferometric imaging results in a greatly improved MIP contrast (Fig. 1b). We constructed a theoretical framework that calculates the defocused interferometric photothermal images of single nanoparticles with different size. This framework provides the optimal MIP detection focus position relative to the nominal focus position of the particles. Compared to reported scanning methods[36, 37, 38], we demonstrate vibrational detection of 100 nm polymethyl methacrylate (PMMA) particles at similar signal-to-noise ratio (SNR) level but with a 3 orders of magnitude higher throughput. By tuning the IR wavenumber, WIDE-MIP spectra of single viruses are acquired from the hyperspectral images (Fig. 1c). We systematically recorded the fingerprints of single vaccinia viruses (VACV), a DNA poxvirus related to Monkeypox[40], single vesicular stomatitis viruses (VSV), the prototype RNA virus[41], and varicella-zoster viruses (VZV), an DNA virus included in the herpesvirus group[42]. Dramatically, the spectra provide signatures of not only viral proteins, but also nucleic acids of individual viruses. Nucleic acid peaks of thymine (T) and uracil (U) residue vibrations in VACV and VSV were detected respectively, indicating unique IR signature of DNA and RNA viruses. Besides the contents, WIDE-MIP data further suggests a β enriched sheet structure in VZV, showing the potential of analyzing protein secondary structure in a single virus.

**Theory and experimental validation of WIDE-MIP detection.**

WIDE-MIP is a highly sensitive vibrational detection platform based on infrared photothermal modulation of interferometric scattering. The schematic of the WIDE-MIP microscope is illustrated in Fig. 1a. In previous implementation of widefield MIP,[39] a visible LED was utilized as the probe light, which had a relatively long pulse duration of ~ 1 μs and only allowed detection of PMMA beads of 1.0 μm diameter. To match the nanosecond-scale thermal decay of nanoparticles (240 ns for 200-nm PMMA beads in air[28]), we incorporated a nanosecond pulsed laser (NPL52C, Thorlabs, pulse duration of 129 ns) as the visible probe to improve the sensitivity. A pulsed mid-infrared laser (Firefly-LW, M Squared Lasers) excites the sample placed on a silicon substrate. The visible probe $E_i$ illuminates the sample and is further scattered by the sample $E_s$ and reflected by the substrate $E_r$. Compared to a transparent substrate, such as calcium fluoride, silicon reflects most of the forward-scattered light and increases the total back-scattering[43]. Consequently, the scattered light is interfered with the reflected light and the resulting interferometric image represents the coherent sum of the scattered and reflected fields[43, 44]:

$$I_{det} = |E_r + E_s|^2 = |E_r|^2 + |E_s|^2 + 2|E_r||E_s|\cos\varphi \qquad (1)$$

where $\varphi$ is the phase difference between $E_s$ and $E_r$. The normalized interferometric contrast $S$ is defined as:

$$S = \frac{I_{det} - I_{bkg}}{I_{bkg}} = \frac{|E_r + E_s|^2 - |E_r|^2}{|E_r|^2} = \frac{|E_s|^2}{|E_r|^2} + 2\frac{|E_s|}{|E_r|}\cos\varphi \qquad (2)$$

where $I_{bkg}$ is the background intensity.

For particles of small size like viruses, $|E_r|^2 \gg |E_s|^2$. Then, we have

$$S \cong 2\frac{|E_s|}{|E_r|}\cos\varphi \qquad (3)$$

The photothermal contrast $C$ induced by IR absorption is generated from the interferometric contrast difference between IR on (hot) and IR off (cold) states:

$$C = \frac{2}{|E_r|}(E_s^{hot}\cos\varphi^{hot} - E_s^{cold}\cos\varphi^{cold}) \qquad (4)$$

where $|E_r|$ is assumed as a constant in the modulation. For the purpose of brevity, only the change of $E_s$ is taken into account between hot and cold states and $\varphi$ is considered as a constant in previous MIP work[38, 45, 46]. However, for specular reflection, $E_r$ only travels in one direction and is reflected back along the optical axis, while $E_s$ travels in all directions, mostly at oblique angles. Due to thermal expansion of the particle, the travelling direction of $E_r$ relative to $E_s$ is different, and thus the phase angle $\varphi$ is slightly different in hot and cold states. Because the phase angle $\varphi$ also depends on the axial position of the optical focus, the MIP contrast can be optimized by tuning the focal position. To precisely control the Z-axis scanning, the objective is mounted on an objective piezo scanner for defocus-enhanced photothermal image acquisition (Fig. 1b).

To validate the interferometric phase difference, the interferometric image of a 200 nm diameter (D) PMMA bead was numerically simulated via the boundary element method (BEM)[47]. Interferometric contrast $S$ is then calculated using the metallic nanoparticle boundary element method (MNPBEM) toolbox[45]. The MIP signal is generated from the interferometric contrast difference between IR on (hot) and IR off (cold) states. The transient temperature difference between hot and cold states is set be ~80 K over a temporal window of 129 nanoseconds (duration of probe pulse), which is calculated from COMSOL simulation[36, 45] (Supplementary Fig. 2, details in Supplementary Note 1). We simulated the interferometric images of the 200 nm PMMA bead at both cold (T = 293.15 K) and hot (T = 373.15 K) states along the Z-axis focus of the objective. The interferometric contrast at the center of the diffraction-limited image of the PMMA bead on a silicon substrate is calculated as the focus position Z is swept. Here, Z is set to be zero for exact optical focusing at the sample-substrate interface for the light-collecting objective. As shown in Fig. 2a, the simulated defocus curves of cold and hot contrasts have a similar sinusoidal function shape, both of them reaching the maximum contrast near Z = 0.4 µm. For the hot state, the increased local temperature changes the opto-physical properties of the PMMA particle, such as size ($r$) and refractive index ($n$). As seen from the zoomed-in view at different focal planes, the slopes of the interferometric contrast vary greatly as a function of Z (Fig. 2, b and c). Strikingly, the interferometric contrast is least sensitive to the axial focus when the particle contrast is maximized at Z = 0.4 µm[15]. Consequently, the photothermal contrast is only 0.00035% at Z = 0.4 µm (Fig. 2c). On the contrary, the interferometric contrast is most sensitive to the change in opto-physical properties of the particle caused by the local temperature increase near the interface at Z = 0 µm, where the photothermal contrast is about 0.6% at $\Delta T$ = 80 K (Fig. 2b). Therefore, the difference between cold and hot state, defined as the MIP contrast, is maximized at a defocused plane relative to the interferometric contrast. The interferometric image shows a bright contrast at the focal plane of Z = 0.4 μm, where the MIP contrast is low. The MIP image reaches its maximum contrast at Z = 0 μm, where the interferometric image shows a negative contrast (Fig. 2, d and e).

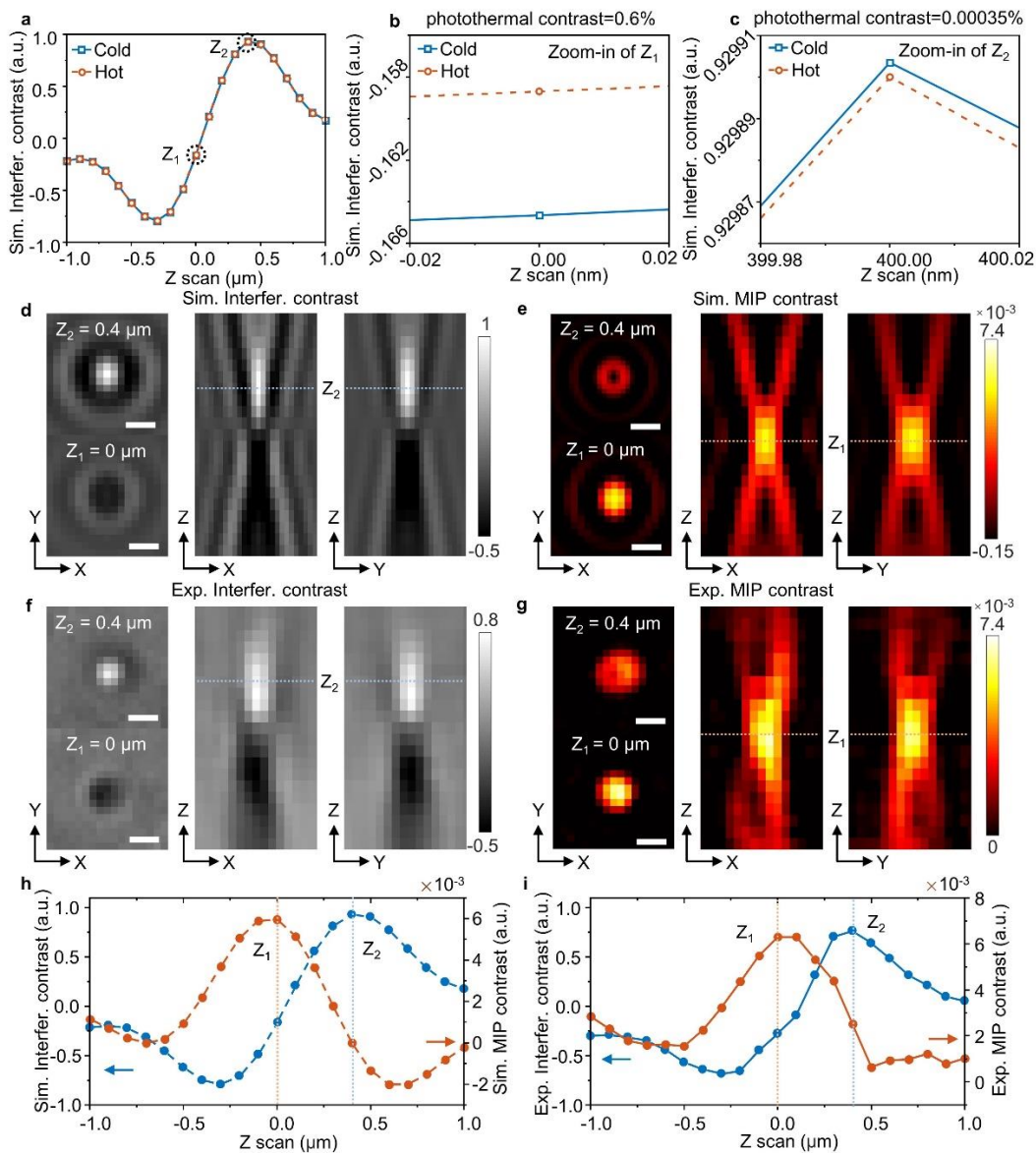

**Fig. 2 Simulation and experimental validation of interferometric defocus-enhanced photothermal contrast.** PMMA beads of D = 200 nm were used as the testbed. (a) Simulated defocus curves of interferometric contrast at the cold (T = 293.15 K) and hot (T = 373.15 K) states. Zoomed-in view of simulated defocus curves of interferometric contrast at the position of (b) $Z_1$ = 0 μm and (c) $Z_2$ = 0.4 μm. Interfer.: Interferometric. The photothermal contrast is 0.6% at $Z_1$ = 0 μm and 0.00035% at $Z_2$ = 0.4 μm. Simulated interferometric images at $Z_2$ = 0.4 μm, $Z_1$ = 0 μm (left) and interferometric scattering along Z axis (right). (e) Simulated MIP images at $Z_2$ = 0.4 μm, $Z_1$ = 0 μm (left) and MIP imaging along Z axis (right). (f) Experimental interferometric images at $Z_2$ = 0.4 μm, $Z_1$ = 0 μm (left) and interferometric scattering along Z axis (right). (g) Experimental MIP images at $Z_2$ = 0.4 μm, $Z_1$ = 0 μm (left) and MIP imaging along Z axis (right). Scale bar: 500 nm. All Z axis images are obtained from Z = -1 to 1 μm. (h) Simulated and (i) experimental defocus curves of interferometric and MIP contrast. Power before the objective: pump: 48 mW at 1728 cm$^{-1}$, probe: ~1 mW. Image acquisition time: 2.36 s per image. Z-axis scanning step: 100 nm.

To experimentally validate the mechanism of interferometric defocus-enhanced MIP, D = 200 nm PMMA beads on a silicon substrate were used. We first used the interferometric contrast (IR off) to locate the beads under the microscope. Once the beads were observed in the focal plane, the IR laser was turned on to 1728 cm$^{-1}$, which corresponds to the acrylate carboxyl vibration (C=O stretching) in PMMA. To optimize the MIP contrast, we manually adjusted the defocus with a piezo scanner. Subsequently, a series of interferometric and MIP images of the PMMA beads were acquired by scanning the focal position of the objective lens. As shown in Fig. 2f and 2g, the experimental images match the simulation results very well.

To derive the optimal condition for MIP imaging, we plotted the interferometric and MIP contrasts as a function of optical focus position through both simulation (Fig. 2h) and experiment (Fig. 2i). The simulated focal plane difference between interferometric and MIP images, $\Delta Z = 400$ nm, is highly consistent with the experimental result. By defocusing the interferometric images, the MIP contrast is increased by 2.5 times for 200 nm PMMA particles. For PMMA beads with different sizes, the defocus curve and $\Delta Z$ have different shapes and values (Supplementary Fig. 3, details in Supplementary Note 2). Thus, this framework provides a guideline to obtain a well-defined and optimized photothermal detection signal by adjusting the focus for MIP detection of nanoparticles with different sizes.

**Hyperspectral performance and spectral fidelity.**

To further test the capability of WIDE-MIP for spectroscopic imaging of single virus in the fingerprint window, we evaluated the hyperspectral performance and spectral fidelity of the system. Single 200-nm PMMA beads with known IR absorption spectrum were chosen for their similar size and dielectric constant (n ≈ 1.5) to the Monkeypox viruses. Fig. 3a and 3b show MIP images of the beads at 1452 cm$^{-1}$ and 1728 cm$^{-1}$, indicating bond-selective contrast from the C−H and C=O stretching of PMMA. The statistical spectra of 30 individual beads showed the distinguished resonance peaks of both C−H and C=O stretching vibrations (Fig. 3c, red line). The standard deviation of the mean MIP contrast within the range of ~1510 to 1610 cm$^{-1}$ was found to be ~ 0.16%, which corresponds to the off-resonance region of PMMA vibration. This demonstrates the stable hyperspectral performance of WIDE-MIP. Furthermore, the spectral fidelity was confirmed by comparing the WIDE-MIP spectrum to FTIR absorption spectrum of PMMA (Fig. 3c, black line)[48]. With the increased MIP contrast, WIDE-MIP realizes the high-speed widefield photothermal detection of D = 100 nm PMMA nanoparticles (Supplementary Fig. 4, details in Supplementary Note 3), which increases the throughput by 3 orders of magnitude compared with scanning MIP at a similar level of SNR (Supplementary Table 1)[36, 37, 38].

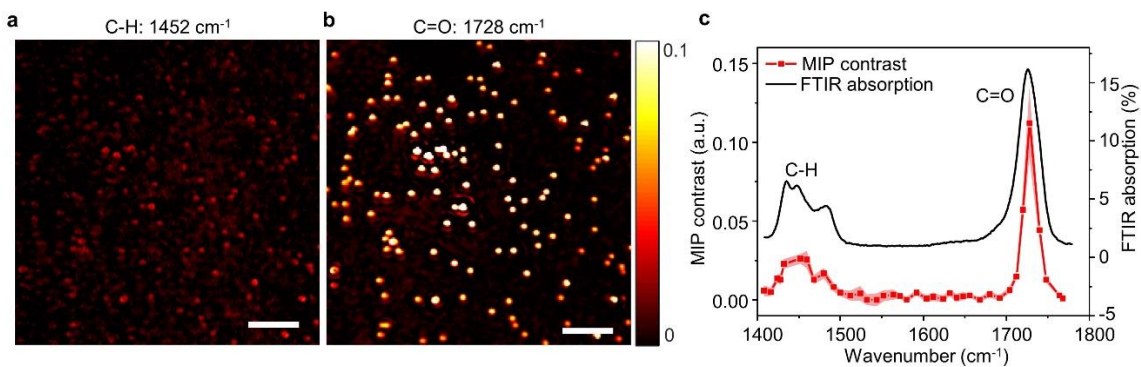

**Fig. 3 Hyperspectral performance and spectral fidelity of WIDE-MIP microscopy.** MIP image of D = 200 nm PMMA beads with IR excitation at (a) 1452 cm$^{-1}$ and (b) 1728 cm$^{-1}$. Scale bars: 5 μm. (c) MIP spectrum (red) and FTIR spectrum (black) of D = 200 nm PMMA beads. Power before the objective: pump: 31.4 mW at 1452 cm$^{-1}$, 38.6 mW at 1728 cm$^{-1}$, probe: ~1 mW. Image acquisition time: 2.36 s per wavenumber. The MIP spectrum was normalized by the IR power. The FTIR spectrum was acquired by an attenuated total reflection FTIR spectrometer.

**Fingerprinting and base residue detection of single DNA and RNA viruses.**

With the high-resolution and high-throughput capability, WIDE-MIP opens the possibility of single-virus chemical detection. We used single VACV and VSV viruses as testbeds. The dimensions of the VACV virion are roughly 360 × 270 × 250 nm[49]. VSV is a bullet-shaped RNA virus with a smaller size of 80 × 180 nm[38, 50, 51]. Fig. 4a shows the defocused interferometric scattering image of single VACV viruses. It should be noted that the depth of focus for MIP imaging is 503 nm and the spatial resolution is 417 nm, measured from a 200 nm PMMA particle in MIP image captured at the defocus plane of Z = 0 μm (Supplementary Fig. 5, details in Supplementary Note 4). To confirm MIP imaging of single viruses, both VACV and VSV viruses were expressed with an enhanced green fluorescent protein (eGFP) envelope for an orthogonal validation. With the good overlay of the widefield fluorescence imaging (Fig. 4b) and the interferometric scattering images (Fig. 4c), we confirmed that the observed particles were indeed VACV virions. Atomic force microscope analysis further confirmed the size of single virions (Supplementary Fig. 6, details in Supplementary Note 5). Bond-selective MIP imaging showed the amide II (1544 cm$^{-1}$) and amide I (1656 cm$^{-1}$) vibrational contrasts contributed by viral proteins (Fig. 4, d and e), whereas the off-resonance images at 1768 cm$^{-1}$ showed no contrasts (Fig. 4f). Similar results of single VSV viruses are shown in the defocused interferometric scattering, fluorescence, and MIP images (Fig. 4, g to l). As VSV is less concentrated on the imaging plate than VACV due to the preparation procedure, we provided more data in Supplementary Fig. 7.

To provide further insight into the viral structure and content, we performed WIDE-MIP hyperspectral imaging of single VACV and VSV viruses (blue and red arrows labeled in Fig. 4, a to l). Obvious differences were overserved in the single-virus fingerprints (Fig. 4m). Merited from the high-throughput ability of spatial multiplexing of WIDE-MIP, spectral analysis of multiple viruses was performed. The statistical spectra of both VACV (n = 36) and VSV (n = 33) (Fig. 4, n and o) are in good agreement with the single-virus spectra (Fig. 5m). Besides the viral protein vibrations, some unique peaks reveal the information of the viral nucleic acids. Different from the wide amide I peak from the pure protein samples (Supplementary Fig. 8a), the strongest sharp peak at 1656 cm$^{-1}$ is contributed by a superposition of the the viral protein, adenine (A) and T residue vibrations in viral DNA of VACV (Fig. 4n). A medium feature at 1580 cm$^{-1}$ is assigned to the T residue vibration in VACV viral DNA. For VSV, the U residue vibration in RNA is indicated by the strong peak at 1640 cm$^{-1}$ (Fig. 4o). More detailed features of nucleic acids in VSV are revealed by the weak peak at 1604 cm$^{-1}$ (A and cytosine (C)) and strong peak at 1656 cm$^{-1}$ (A and proteins). The guanine (G) residue vibrations are identified at 1692 cm$^{-1}$ in both VACV and VSV. The assignments of the chemical components were validated by the pure protein, DNA and RNA film samples[52] (Supplementary Fig. 8, details in Supplementary Note 6). It indicates that WIDE-MIP can provide rich chemical content information of viral proteins and even nucleic acids inside a single virus.

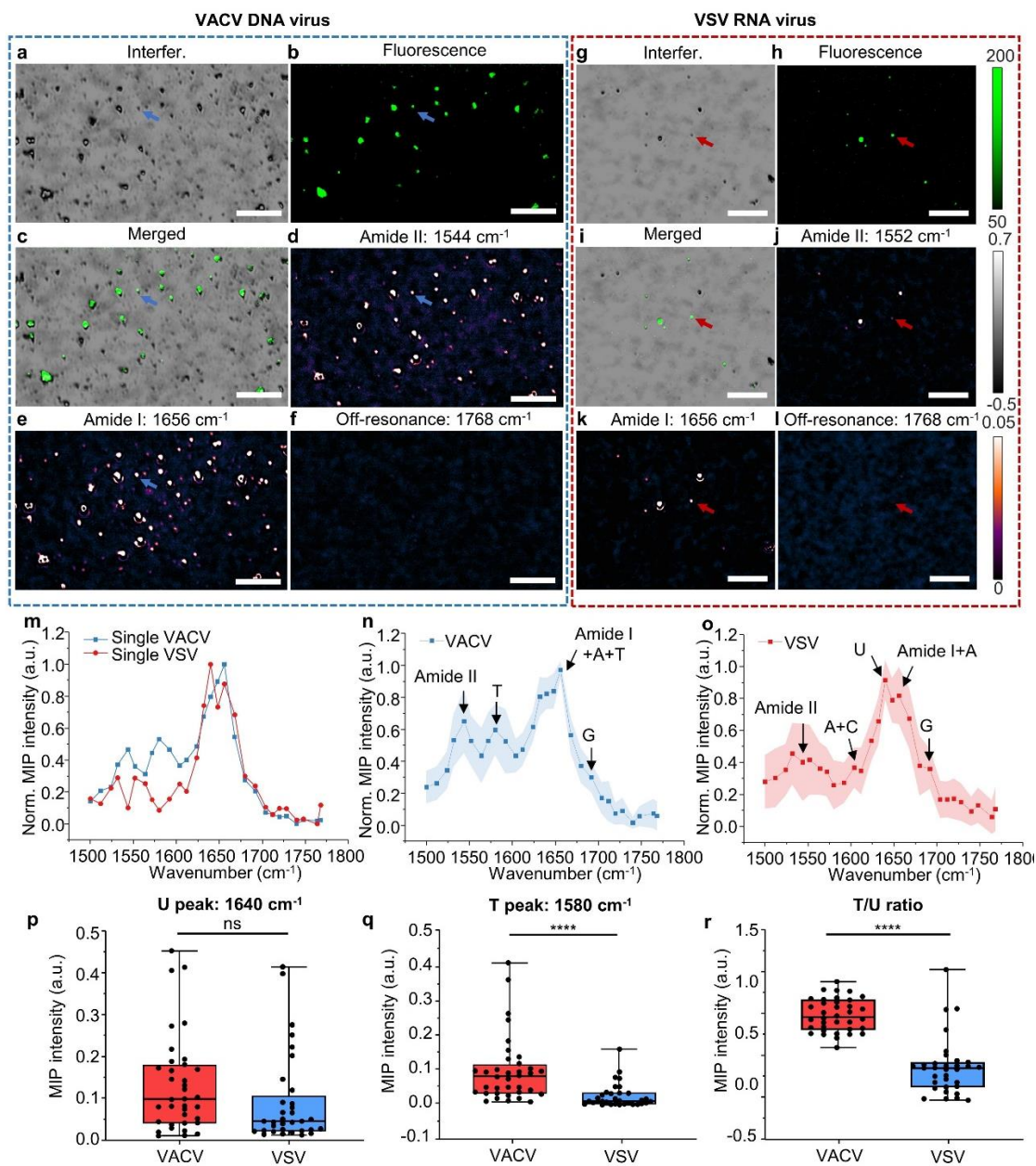

**Fig. 4 Fingerprinting detection of single VACV and VSV viruses.** (a) Defocused interferometric scattering, (b) fluorescence and (c) merged images of single VACV viruses. (d) Amide II bond-selective image of the same area with the pump at 1544 cm$^{-1}$. (e) Amide I bond-selective image of the same area with the pump at 1656 cm$^{-1}$. (f) Off-resonance image showed no contrast. (g) Defocused interferometric scattering, (h) fluorescence and (i) merged images of single VSV viruses. (j) Amide II bond-selective image of the same area with the pump at 1552 cm$^{-1}$. (k) Amide I bond-selective image of the same area with the pump at 1656 cm$^{-1}$. (l) Off-resonance image showed no contrast. Scale bars: 10 μm. (m) MIP spectra of two single VACV and VSV viruses (blue and red arrows labeled). Statistical MIP spectra obtained from (n) 36 single VACV and (o) 33 VSV viruses. Power before the objective: pump: 22.9 mW at 1544 cm$^{-1}$, 29.1 mW at 1552 cm$^{-1}$, 34.5 mW at 1656 cm$^{-1}$, 35.8 mW at 1768 cm$^{-1}$, probe: ~1 mW. Image acquisition time: 2.36 s per wavenumber. The MIP spectrum is normalized by the IR power. Quantified MIP contrast of peaks at (p) T residue and (q) U residue of VACV and VSV. (r) Quantified MIP contrast ratio of peaks at T residue/U residue

of VACV and VSV. Line, median; box, SD. Ns (P ≥ 0.05) denotes no statistically significant difference. Asterisks **** (P < 0.0001) denotes statistically significant difference.

To further demonstrate the potential of WIDE-MIP to differentiate single RNA viruses from single DNA viruses, we compared the signature peaks of nucleic acids at the single-virus level by quantifying the MIP contrast of peaks at T residue and U residue for VACV and VSV to highlight the spectroscopic difference of DNA and RNA viruses. Although the MIP contrasts of VACV and VSV shows no statistically significant difference at 1580 cm$^{-1}$ representing U residue (Fig. 4p, P = 0.138892), MIP contrasts of both T residue (Fig. 4q, P = 0.000074) and the ratio of T/U (Fig. 4r, P < 0.000001) show significant difference between VACV and VSV. Our results show that fingerprint WIDE-MIP has the potential to rapidly classify RNA and DNA viruses in clinic by bond-selective imaging of T and U residues.

**Identification of protein secondary structure in a single virus.**

As the profile in the amide I band is very sensitive to the protein secondary structures[53], WIDE-MIP can be a promising tool to characterize the protein structures compared to expensive and time-consuming approaches such as electron microscopy[54, 55]. To explore such potential, we acquired the WIDE-MIP spectra of another DNA virus, VZV, shown in Fig. 5. It is reported that there are three envelope proteins, glycoprotein B, glycoprotein H and glycoprotein L serving as the most essential VZV proteins that function as the core fusion complex[56]. These proteins have known 3D structure and all of them have a big proportion of β-sheet, and the proportion of turn cannot be ignored[57]. Fig. 5a to c show the defocused interferometric scattering image and bond-selective MIP image of amide II and amide I vibrations of single VZV viruses. Although there are some aggregates in the interferometric image, a lot of single virions are shown. Zoomed-in views of four single VZV virus are illustrated in Fig. 5, d to i. WIDE-MIP spectra of these four single VZV (red arrows labeled in Fig. 5, d to i) were further obtained (Fig. 5j). The specific IR peaks of DNA virus were observed at 1580 cm$^{-1}$ and 1612 cm$^{-1}$, indicating the vibrations of A, C and T residues in the viral DNA of VZV. Compared to the spectra of VSV and VACV, two broad peaks are observed at around 1630-1640 cm$^{-1}$ and 1668 cm$^{-1}$, which are assigned to the β-sheet and the turn structure in the viral proteins of VZV, respectively [53]. The statistical spectra acquired from 30 VZV virions (Fig. 5k) further reveals an enriched β-sheet protein components, viral DNA and lipids in VZV[58] (details in Supplementary Note 6). The spectral fidelity was confirmed by FTIR absorption spectrum of VZV powder (Fig. 5l). Therefore, besides the detection of major chemical components, WIDE-MIP can identify the protein secondary structure related to their function in a virus.

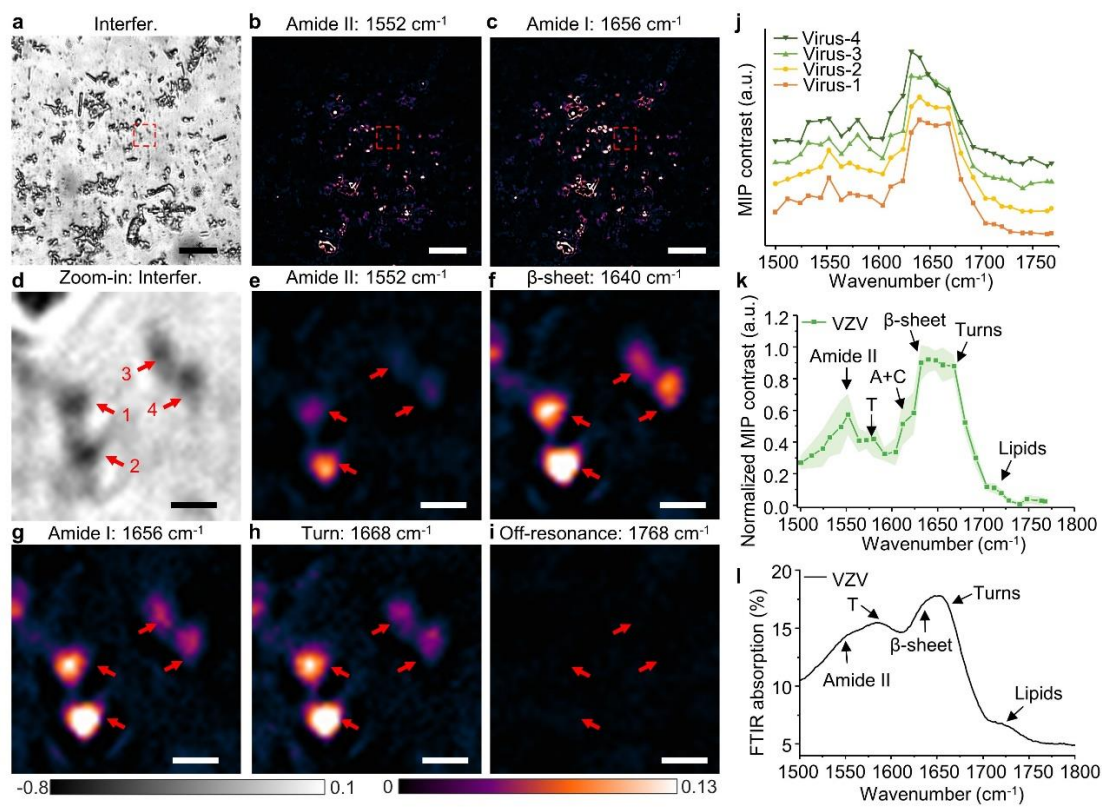

**Fig. 5 Protein secondary structure identification in single VZV viruses.** (a) Defocused interferometric scattering, (b) Amide II and (c) Amide I bond-selective image of single VZV viruses with the pump at 1552 cm$^{-1}$ and 1656 cm$^{-1}$, respectively. Scale bars: 10 μm. (d) Zoomed-in view of defocused interferometric scattering of four viruses in (a). (e) Amide II, (f) β-sheet, (g) Amide I, (h) Turn bond-selective image of the same area in (d) with the pump at 1552 cm$^{-1}$, 1640 cm$^{-1}$, 1656 cm$^{-1}$, 1668 cm$^{-1}$, respectively. (i) Off-resonance image showed no contrast of the same area in (d) with the pump at 1768 cm$^{-1}$. Scale bars: 1 μm. (j) MIP spectra of four VZV viruses in (d) (red arrows labeled). (k) Statistical MIP spectra obtained from 30 single VZV viruses. Power before the objective: pump: 29.1 mW at 1552 cm$^{-1}$, 33.1 mW at 1640 cm$^{-1}$, 34.5 mW at 1656 cm$^{-1}$, 34.1 mW at 1668 cm$^{-1}$, 35.8 mW at 1768 cm$^{-1}$, probe: ~1 mW. Image acquisition time: 2.36 s per wavenumber. The MIP spectrum is normalized by the IR power. (l) FTIR spectrum of pure VZV virus powder. The FTIR spectrum was acquired by an attenuated total reflection FTIR spectrometer.

## Discussion

We present a single virus fingerprinting approach, termed WIDE-MIP microscopy, that addresses the unmet need for identification of single virus. Our method allows composition detection of viral nucleic acids and proteins with high-throughput. A theoretical framework for interferometric defocus-enhanced photothermal signal is developed and experimentally validated, providing a guideline to obtain a well-defined photothermal signal by adjusting the defocusing. Compared to scanning MIP, WIDE-MIP increases the imaging throughput by 3 orders of magnitude for fingerprint analysis of nanoparticles at the same SNR. Besides content detection of viral proteins, viral DNA and viral RNA, WIDE-MIP further identifies the protein secondary structure in single virus by revealing enriched β sheet components.

While we utilized defocused interferometric imaging to improve MIP contrast, our analysis of biological samples relies on MIP contrasts, whereas the interferometric images are solely used for sample localization before acquiring the MIP images. Despite the slight defocusing in the interferometric images, the MIP images remain in focus, revealing distinct contrasts of the biological samples. The MIP images not only provide valuable morphological information of the biological samples, but also offer insightful biochemical details about their specific contents.

In this work, both VACV and VSV viruses were expressed with an eGFP envelope. The eGFP was fused to the VSV G protein, where each VSV contains ~ 1,200 molecules of the G protein on the viral surface[59]. With the formed G protein and G-eGFP fusion protein heterodimers, there are ~ 600 eGFP molecules on the surface of a single virus. Comparing the size of eGFP (2.4 × 4.2 nm) to that of the VSV virus (80 × 180 nm), we estimate that only 1.3% of a single VSV virus consists of eGFP. Thus, the effect of eGFP on MIP imaging should be negligible due to the relatively low content of eGFP in a single virus.

Supporting this notion, eGFP has a β-enriched sheet structure, while no obvious β sheet chemical signature was observed in either VACV or VSV expressing eGFP in Fig. 4. Additionally, the pure VZV viruses without any labeling showed an enriched β sheet component. These findings further support that the effect of eGFP on MIP imaging is minimal. Moreover, we note that the key distinction between DNA and RNA viruses lies in the different nucleic acid peaks associated with T and U residue vibrations, which is also unrelated to the eGFP proteins.

For the analysis of actual virus samples, label-free methods may be more suitable for diagnostic purposes. Thus, in this work, we firstly performed fluorescence-guided MIP analysis of single viruses by integrating fluorescence imaging and MIP imaging for accurate virus identification via WIDE-MIP (Fig. 4). Subsequently, we performed MIP imaging and obtained fingerprint spectra of unlabeled pure VZV viruses to achieve label-free detection of single viruses (Fig.5). These approaches allow for comprehensive analysis while minimizing any potential influence of fluorescence labeling on the MIP imaging results.

In comparison to recently reported fluorescence-detected MIP (F-MIP) microscopy[60, 61], WIDE-MIP offers a distinct advantage in detecting bionanoparticles with low levels of expressed fluorescence tags (Supplementary Fig. 9, details in Supplementary Note 7). Although the photobleaching of aggregated eGFP-VACVs showed a similar level in F-MIP[60] (< ~ 10%), severe photobleaching was observed in single VACVs (> ~ 95%). This photobleaching of single viruses limits the detection of photothermal modulation and acquisition speed. Considering that the MIP signal relies on the difference in fluorescence intensity between the IR-on (hot) and IR-off (cold) states, this severe bleaching at the single-virus level further compromises the reliability of F-MIP. Instead, we focused on fluorescence-guided WIDE-MIP analysis, which enables label-free chemical imaging of single viruses.

Benefit from the compositional analysis of single viruses in a label-free manner, we envision WIDE-MIP as an alternative analysis tool for viral vectors used in gene therapy. Viral vectors, including adeno-associated viruses, adenoviruses, and lentiviruses, are increasingly used in gene therapy but pose challenges for quality control testing and characterization due to their complexity[62, 63]. To ensure a safe, consistent, and high-quality product, accurate and rapid analytical assays are needed to monitor quality attributes. Sodium dodecyl-sulfate polyacrylamide gel electrophoresis, mass spectrometry, immunoblotting, enzyme-linked immunosorbent assay, polymerase chain reaction, or transmission electron microscopy are used to identify protein, genome and capsid content[64, 65], but these assays can be time-consuming and require pre-treatments or extraction. To address these limitations, we further demonstrated high-speed chemical imaging of single VACV by reducing the acquisition time to 0.32 s per image per wavenumber of single viruses and the SNR of one single VACV is ~ 4 within the field of view of 24 by 24 μm (Supplementary Fig. 10). With the ability to rapidly acquire fingerprints of single viruses, WIDE-MIP can provide insights into the quality control of viral vectors, such as their identity, purity, and stability[64] (details in Supplementary Note 8).

Future improvement for WIDE-MIP can focus on fingerprinting viruses or exosomes in liquid conditions, allowing for detecting biological nanoparticle samples in their natural states.

This can be achieved by incorporating microfluidic systems[18] and designed substrates[43] to capture viruses and enhance imaging contrast in liquid measurements, which will broaden the applicability of WIDE-MIP to real-world applications.

## Methods

**Materials.** The double-side polished silicon wafer (4 inch, 500 μm thickness) was purchased from University Wafer and diced into 10 mm × 20 mm pieces. PMMA nanoparticles were purchased from Phosphorex. 0.1% poly-l-lysine and bovine serum albumin (BSA) were purchased from Sigma-Aldrich. Inactivated varicella zoster virus strain VZ-10 was purchased from Fisher Scientific.

**Lab-built WIDE-MIP microscope.** The IR pump beam was generated by a tunable (from 1400 to 1800 cm$^{-1}$) mid-IR laser (Firefly-LW, M Squared Lasers) operating at 20 kHz repetition rate with a ~20 ns pulse duration. The pump pulses were modulated by an optical chopper (MC2000B, Thorlabs). For the wide-field photothermal imaging, the IR beam was weakly focused on the sample plane from the bottom of the silicon substrate via an off-axis parabolic mirror. A delay pulse generator (9254, Quantum Composers) was used to synchronize the pump pulse, the probe pulse and the interferometric pulse recorded by the camera. For the power normalization, a power meter (PM16-401, Thorlabs) was used to monitor the IR power. The visible probe was provided with a pulsed 520 nm nanosecond laser (NPL52C, Thorlabs) with a pulse duration of 129 ns. The probe laser illuminated the sample from the top through a 50/50 polarizing beam splitter, a quarter-wave plate and a high numerical aperture (NA) air objective (MPLFLN Olympus, 100X, NA 0.9). To acquire the defocus-enhanced photothermal images, the objective was adapted with an objective piezo scanner (Piezosystemjena, MIPOS 100), which can provide precise Z-axis scanning in steps of 100 nm. The incident light was then scattered by the sample and reflected by the silicon substrate. The consequent interferometric scattering was collected by the same objective and recorded by a complementary metal-oxide semiconductor (CMOS) camera (Q-2HFW, Adimec). We further employed a 2 million well-depth camera to receive sufficient probe photons at each pixel.

**Theoretical simulation.** A theoretical framework was developed to calculate the focus-dependent interferometric and photothermal images of nanoparticles of different sizes. An image field representation of optical fields was employed by considering imaging optics and system parameters. The simulation was built upon the previously developed model via the boundary element method (BEM), which is a computationally efficient approach for calculating the interferometric scattering from a nanoparticle near a substrate. A custom-developed metallic nanoparticle boundary element method (MNPBEM) toolbox was used to solve Maxwell's equations for a dielectric environment where the nanoparticles have homogeneous and isotropic dielectric functions and are separated by abrupt medium interfaces [43]. The MNPBEM was implemented in MATLAB and the simulation could be compartmentalized into five steps: (1) dielectric functions initialization of nanoparticle, substrate and environment to define the system geometry, such as the radius ($r$) and refractive index ($n$) of the nanoparticle; (2) specification of excitation scheme, such as incident illumination wavelength ($\lambda$) and illumination function; (3) solver setup for the BEM equations; (4) BEM equations' solutions for the given excitation; (5) calculation of the far-field scattered field and image fields of the nanoparticle. We assumed that a PMMA particle ($r$, $n$ = 1.49 [66]) was placed on top of a silicon substrate ($n$ = 4.2 [67]). The interferometric scattered field was calculated as the total backscattered field considering the reflections from the silicon surface using Green's functions. The image fields were then simulated via angular spectrum representation integral and the detected interferometric signals were calculated according to equation (2) in the main text. The photothermal signals were then generated from the interferometric scattering difference between IR on (hot) and IR off (cold) states. We simulated the interferometric images of a PMMA bead with different sizes at both cold (T=293.15 K) and hot (T=373.15 K) states along the Z-axis scanning of objective. The interferometric and photothermal contrasts were recorded at the center of the diffraction-limited image of the PMMA bead as the focus position Z sweeping. Here, Z was set to be zero for exact optical focusing at the sample-substrate interface for the light-collecting objective, where the numerical aperture of the objective was also considered for the collecting angle. Thus, the defocus curves of both interferometric and photothermal contrasts were obtained.

**Fluorescence imaging of single virus.** A 488 nm diode laser (200 mW, Cobolt 06-MLD) was used for fluorescence excitation. The excitation beam was expanded through a 4-f system ($f_1$ = 50 mm, LA1131-A-ML and $f_2$ = 300 mm, LA1484-A-ML, Thorlabs) and then coupled into the light path of probe laser. The fluorescence emission went through the same objective lens and was collected with a filter set (Excitation filter: FES0500, Thorlabs; Dichroic beam splitter: Di03-R405/488/532/635-t1-25x36, Sermock; Emission filter: FF01-525/30-25, Sermock). A CMOS camera (FLIR, Grasshopper3GS3-U3-51S5M) was used to capture the fluorescence images and the exposure time was set to 5 s for optimized contrast. Virus samples on silicon substrate were first imaged by fluorescence to confirm the single virus and then imaged with WIDE-MIP at the same position.

**Data processing.** The interferometric and MIP images were acquired using a lab-built Labview program and analyzed with ImageJ, detailed methods were described in previous work [39]. The interferometric images were captured at a camera shutter speed of 1270 Hz. The MIP images were obtained as the intensity difference between the hot and the sequential cold frame, at the speed of 635 frames/s. The interferometric images were normalized by the background reflection. Pseudocolor was added to the MIP and fluorescent images with Image J software or MATLAB. The SNR was calculated from the ratio between the mean value of the center region (25 pixels) of the single particle in resonance photothermal image and the standard deviation of the off-resonance photothermal.

**Sample preparation.** The silicon wafers were cleaned in sequence with acetone, ethanol, and deionized (DI) water rinse. For PMMA nanoparticle detection, the PMMA beads were diluted ~100 times with DI water and then spin-coated on the silicon substrate and dried in air. For virus analysis, the VACV and vesicular stomatitis virus (VSV) samples were prepared according to the previous method [38]. Both the recombinant VSV expressed an enhanced green fluorescent protein (eGFP) and VACV expressed Venus. To load viruses onto the substrate, the silicon surface was incubated with 0.1% poly-l-lysine for 1 h. Then, 100 μL of either VACV or VSV stock was incubated in the center of each poly-L-lysine coated silicon for 1 h at room temperature. Both VACVs and VSVs were diluted to ~1×10⁸ PFU/mL. All virions were crosslinked and inactivated using 1.0 mL of 4% formaldehyde for 1 h. After modification, the substrate was rinsed with sterile filtered DI water and then dried in air. For varicella-zoster virus (VZV) detection, the lyophilized VZV pellet was dissolved in 200 μL PBS and filtered by a 0.22 μm filter. Then, 100 μL of VZV stock was incubated in the center of each poly-L-lysine coated silicon for 1 h at room temperature. After modification, the substrate was rinsed with sterile filtered DI water and then dried in air. To fabricate the pure protein film sample, 10 μL 10 mg/mL BSA solution was dropped onto the silicon surface and dried in air. The pure DNA and RNA solutions were prepared from the cDNA of melanoma cell and ssRNA of T24 cell, respectively, as described earlier [68]. The pure DNA and RNA films were prepared by dropping 5 μL cDNA and ssRNA onto the silicon surface, repectively, and dried in air.

**FTIR measurement.** The FTIR spectra of 200-nm dry PMMA beads and viruses were measured on an attenuated total reflection FTIR spectrometer (Nicolet Nexus 670, Thermo Fisher Scientific). The spectra resolution is 2 cm$^{-1}$ and each spectrum was measured with 128 scanning. All spectra were automatically normalized by the baseline correction on the system.

**Data availability**

All data supporting the findings of this study are available within the article and the Supplementary Information. Source data are provided with this paper.

**Code Availability**

MATLAB codes for simulation are available in Zenodo (DOI: 10.5281/zenodo.7957859).

**References**

_1. Farahat RA, Abdelaal A, Shah J, Ghozy S, Sah R, Bonilla-Aldana DK, *et al.* Monkeypox outbreaks during COVID-19 pandemic: Are we looking at an independent phenomenon or an overlapping pandemic? *Ann. Clin. Microbiol. Antimicrob.* **21**, 26 (2022).
2. Lai CC, Hsu CK, Yen MY, Lee PI, Ko WC, Hsueh PR. Monkeypox: An emerging global threat during the COVID-19 pandemic. *J. Microbiol. Immunol. Infect.* **55**, 787-794 (2022).
3. Mahase E. Seven Monkeypox cases are confirmed in England. *BMJ* **377**, o1239 (2022).
4. Weissleder R, Lee H, Ko J, Pittet MJ. COVID-19 diagnostics in context. *Sci. Transl. Med.* **12**, eabc1931 (2020).
5. Smyrlaki I, Ekman M, Lentini A, Rufino de Sousa N, Papanicolaou N, Vondracek M, *et al.* Massive and rapid COVID-19 testing is feasible by extraction-free SARS-CoV-2 RT-PCR. *Nat. Commun.* **11**, 4812 (2020).
19

**Acknowledgments**

This research was supported by R35GM136223 to JXC. Authors thank Fukai Chen for providing the pure DNA and RNA samples.

**Author contributions**

Q.X. and J.-X.C. designed the experiments. Q.X. performed the experiments and analyzed the data. Z.Y.G., H.N.Z. and C.Y. helped in data analysis. Z.Y.G. and Q.X. constructed the setup. Z.Y.G. coded the program for data acquisition. C.Y., H.N.Z. and M.S.U. provided guidance and discussions on the simulation. S.S. prepared the VACV and VSV samples and provided constructive suggestions in virus analysis. L.W. performed the atomic force microscope analysis. Q.X. wrote the manuscript with contributions from J.-X.C., Z.Y.G., C.Y. and H.N.Z. J.-X.C. and J.H.C. provided overall guidance for the project. All authors have given approval to the final version of the manuscript.

**Competing interests**

The authors declare no competing interests.

Supporting Information for

# Single virus fingerprinting by widefield interferometric defocus-enhanced mid-infrared photothermal microscopy

Qing Xia[1], Zhongyue Guo[2], Haonan Zong[1], Scott Seitz[3], Celalettin Yurdakul[1], M. Selim Ünlü[1], Le Wang[1], John H. Connor[3,*] and Ji-Xin Cheng[1,2,4,*]

[1] Department of Electrical and Computer Engineering, Boston University, Boston, Massachusetts 02215, United States;
[2] Department of Biomedical Engineering, Boston University, Boston, Massachusetts 02215, United States;
[3] Department of Microbiology and National Infectious Diseases Laboratories, Boston University School of Medicine, Boston, Massachusetts 02118, United States;
[4] Photonics Center, Boston University, Boston, Massachusetts 02215, United States.

Summary

Number of Pages: 16

**Supplementary Note 1. Simulation of the temperature rise.**

To better understand the photothermal process, a theoretical model was built to solve the temperature difference between hot and cold states. This model was developed using COMSOL based on our previous work[1]. The time-dependent thermal diffusion process can be simulated via the heat-transfer-in-solids module in COMSOL Multiphysics[2]. To calculate the heat dissipation, a heat source term $\boldsymbol{Q(t)}$ is defined as below:

$$C_p \rho \frac{\partial T}{\partial t} + \nabla \cdot (-k \nabla T) = Q(t) \tag{1}$$

where $T$ is the temperature, $t$ is the time, $C_p$ is the heat capacity, $\rho$ is the density, and $k$ is the thermal conductivity of the material in the system.

In this simulation, a 200 nm PMMA bead was sitting on the silicon substrate in air. The IR heating beam size is 24.6 µm by 30 µm and the power is 40 mW measured from the experiments. Both the initial temperature and the simulation boundary were assumed to be 298 K. The heat source was defined as the domain of the PMMA bead. Heat convection was not considered in this simulation. By solving the equation (1), the temperature rise distribution of the bead under single IR pulse heating was simulated via COMSOL 6.0. Supplementary Fig. 2a shows the temperature profile of the system. The calculated $\Delta T$ on the single bead is ~80 K, integrated from the pulse width of single probe pulse, which is ~129 ns (Supplementary Fig. 2b).

**Supplementary Note 2. Interferometric defocus-enhanced MIP signal for PMMA beads with different sizes.**

For PMMA beads with different sizes, the defocus curves of interferometric and MIP contrasts have different shapes. The resulting ΔZ between the maximum interferometric and MIP contrasts varies with the particle size. Similar mechanism of interferometric defocus-enhanced MIP signal was also validated for D = 100 nm and D = 150 nm PMMA (Supplementary Fig. 3, a and b), where the simulated focal plane difference ΔZ is found to be 200 nm and 300 nm, respectively. While for big nanoparticles with diameter of 350 nm, 400 nm and 500 nm, the increased particle size results in a noticeable change of the defocus curve in interferometric imaging[3] (Supplementary Fig. 3, c to e), and the resulted ΔZ was estimated to be < 100 nm. The simulated interferometric and MIP images for D = 500 nm PMMA show similar maximum at Z = 400 and 500 nm, and decreased contrasts at Z = 0 nm (Supplementary Fig. 3f). The experimental defocus curve and images for D = 500 nm PMMA also match the simulation results very well (Supplementary Fig. 3, g to h). It indicates that defocusing does not help MIP enhancement for big particles with diameter larger than 300 nm. Since most viruses vary in diameter from 20 nm to 300 nm, defocusing especially works well for small nanoparticles with similar size of single virus, which provides WIDE-MIP as a promising tool for single virus analysis.

**Supplementary Note 3. Detection limit of WIDE-MIP imaging.**

To demonstrate high-speed widefield photothermal detection of small nanoparticles, we performed WIDE-MIP imaging of both D = 200 nm and D = 100 nm PMMA beads. The defocused interferometric images of beads in air are shown in Supplementary Fig. 4a (D = 200 nm PMMA) and Supplementary Fig. 4b (D = 100 nm PMMA). WIDE-MIP images of both types of PMMA beads showed high contrast at 1728 cm$^{-1}$, indicating C=O vibration in PMMA (Supplementary Fig. 4, c and d), while no contrasts were observed at the off-resonance 1800 cm$^{-1}$ (Supplementary Fig. 4, e and f). The WIDE-MIP images were acquired with the signal averaged for 2.36 s, showing good SNR of ~ 87 for 200 nm PMMA and SNR of ~14 for 100 nm PMMA.

**Supplementary Note 4. Spatial resolution of WIDE-MIP imaging.**

To evaluate the spatial resolution of WIDE-MIP, we first measured the interferometric contrast profile across one single 200 nm PMMA bead at different focal planes. The Gaussian fitted full width at half maximum (FWHM) along the horizontal axes is 278 nm at the interferometric focus Z = 0.4 µm (Supplementary Fig. 5a), which is consistent with the theoretical resolution of the interferometric imaging system. In WIDE-MIP, the interferometric image is defocused at Z = 0 µm, the FWHM along the particles is 404 nm, which is a little larger than the resolution of the interferometric system (Supplementary Fig. 5b). Thus, the spatial resolution is 417 nm, measured as FWHM of particles in the WIDE-MIP image captured at the defocus plane of Z = 0 µm (Supplementary Fig. 5c). In addition, WIDE-MIP system allows depth resolved measurement, the depth of focus for the imaging system is 503 nm, measured as FWHM of the MIP contrast profile in WIDE-MIP image from Z = -1 μm to Z = 1 μm (Supplementary Fig. 5d).

**Supplementary Note 5. Atomic force microscope analysis of single viruses.**

Size and morphology characterizations of single vaccinia viruses (VACV) were performed on an atomic force microscope (AFM) (NanoWizard 4 XP, Bruker Nano) in tapping mode (Supplementary Fig. 6a). AFM analysis further confirmed the single VACV particle 1 on the silicon substrate with a brick-shaped size of 280 × 350 nm (Supplementary Fig. 6b) that is consistent with the reported viral shape and dimensions [4, 5]. VACV particle 2 looks like a denatured rather than intact virion (Supplementary Fig. 6c). It is highly likely happened during the fixation process.

**Supplementary Note 6. Peak assignments of viral WIDE-MIP spectra.**

The assignments of the chemical components were validated by the pure protein, DNA and RNA film samples (Supplementary Fig. 8). The dominant two peaks in the spectrum of pure BSA are contributed by the amide I band at 1650 $cm^{-1}$ and the amide II band at 1550 $cm^{-1}$, indicating the vibrations in the protein[6]. For the assignments of base residues in the nucleic acids, according to the literature[7], the peaks at 1656 $cm^{-1}$ and 1604 $cm^{-1}$ are assigned to the $NH_2$ bending and C=N streching vibrations in adenine (A) residue. The peaks at 1680 $cm^{-1}$ and 1640 $cm^{-1}$ are assigned to the NH in-plane deformation vibration and C=O and C=C stretching vibrations in uracil (U) residue, respectively. The peaks at 1656 $cm^{-1}$ and 1580 $cm^{-1}$ are assigned to the $C_4$=O stretching vibration and ring stretching vibration in thymine (T) residue, respectively. The peak at 1604 $cm^{-1}$ is assigned to the In-plane ring vibrations in cytosine (C) residue. The peak at 1692 $cm^{-1}$ is assigned to the C=O stretching vibration and $NH_2$ scissoring vibration in guanine (G) residue. Although the peak assignments of base residues in literature show ~ 10 $cm^{-1}$ higher than our results, we attribute this small difference to the different sample state, testing environment and instruments. The peak at ~ 1725 $cm^{-1}$ is assigned to the C=O stretching vibration in viral lipids[8].

To further demonstrate the accuracy of MIP spectra for virus fingerprinting, we provided the biochemical components and the FTIR spectrum of pure VZV powder for comparison (Fig. 5l). For a VZV, it has a lipid-rich envelope derived from cellular membranes, within which viral glycoproteins are inserted[9]. Within the VZV, there are ~ 125-kb linear double-stranded DNA genome and ~ 3000 proteins[10]. Notably, three envelope proteins, namely glycoprotein B, glycoprotein H, and glycoprotein L, have been identified as essential VZV proteins forming the core fusion complex[9]. These proteins have known 3D structure, and a significant proportion of their secondary structures consist of β-sheets, with turn structures also present[11]. The results in Fig. 5 showed that WIDE-MIP can accurately identify VZV viruses and reveal T residue vibrations in viral DNA, lipids and enriched β sheet components in VZV viral proteins, which are consistent with the biochemical components in VZV and FTIR spectrum of VZV powder.

**Supplementary Note 7. Comparison between fluorescence-detected mid-infrared photothermal microscopy (F-MIP) and WIDE-MIP.**

While F-MIP imaging offers a larger modulation depth[12, 13], WIDE-MIP offers a distinct advantage in detecting bionanoparticles that have very low levels of expressed fluorescence tags.

One of the key requirements for F-MIP is a robust and high-quality fluorescence signal. In this regard, it is essential to highlight the difference in fluorescence probe labeling quantity between our previous study conducted by Yi Zhang et al.[12] and the current work. In Yi Zhang et al's work, high concentrations of commercial chemical dyes (10 μM Nile Red or Rhodamine) were used for labeling high-content biological components, such as proteins in cells. While in this work, both VACV and VSV viruses were expressed with an enhanced green fluorescent protein (eGFP) envelope. The expressed eGFP was fused to the VSV G protein, where each VSV contains ~ 1,200 molecules of the G protein on the viral surface[14]. With the formed G protein and G-eGFP fusion protein heterodimers, there are ~ 600 eGFP molecules on the surface of a single virus. Comparing the size of eGFP (2.4 × 4.2 nm) to that of the VSV virus (80 × 180 nm), we estimate that only 1.3% of a single VSV virus consists of eGFP. Thus, due to the lower content of total fluorescence probes in this study, the resulting fluorescence intensity is significantly weaker compared to Yi Zhang et al's work. Consequently, the reduced fluorescence intensity poses a challenge when attempting to implement fluorescence-based MIP imaging of single viruses in this work.

To demonstrate it, we used the same camera (FLIR, Grasshopper3GS3-U3-51S5M) as in Yi Zhang et al's work to perform the fluorescence imaging of VACVs expressing eGFP, and tried the same paraments with a camera exposure time of 50 ms and a gain of 20 dB. However, these settings failed to capture the fluorescence of single viruses. We increased the exposure time to 500 ms for imaging aggregated viruses, while using the exposure time of ~ 5 s and maximum gain setting for single viruses (Supplementary Fig. 9a to d). Although the photobleaching of aggregated virus showed a similar level as Yi Zhang et al's work (< ~ 10%, Supplementary Fig. 9e), severe photobleaching was observed in single viruses (> ~ 95%, Supplementary Fig. 9f). This photobleaching of single viruses limits the detection of photothermal modulation and acquisition speed. Considering that the MIP signal relies on the difference in fluorescence intensity between the IR-on (hot) and IR-off (cold) states, this severe bleaching at the single-virus level further compromises the reliability of F-MIP.

On the other hand, scattering-based imaging offers certain advantages, particularly in terms of the photon budget. When compared to fluorescence imaging, scattering-based techniques allow for shorter camera exposure times, higher full well capacity, and reduced saturation issues. In Yi Zhang et al's work, a CMOS camera (FLIR, Grasshopper3GS3-U3-51S5M) with a full well depth of ~ 10,000 and a frame rate of 20 Hz was used for wide-field FMIP imaging. In this work, we employed a camera with a frame rate of 1270 Hz and a full well capacity of 2 million wells (Q-2HFW, Adimec). This choice ensured that sufficient probe photons were received at each pixel, enhancing the quality of the scattering signal. Additionally, the interferometric geometry further enhanced the weak scattering signal of single viruses.

Taking into account these limitations and technical considerations, F-MIP imaging is not suitable for the detection of single viruses. Instead, we focused on fluorescence-guided MIP analysis of single viruses. In this work, we first collected and analyzed the fingerprint spectra of eGFP-virus samples (VACV and VSV) and performed co-localization of fluorescence imaging and MIP imaging to demonstrate the accurate identification of viruses using the WIDE-MIP technique (Figure 4). However, for the analysis of actual virus samples, label-free methods may be more suitable for diagnostic purposes. Thus, we further performed MIP imaging and fingerprint spectra of unlabeled pure VZV viruses to achieve label-free detection of single viruses (Figure 5).

**Supplementary Note 8. Potential application for rapid quality control of viral vectors.**

We further demonstrated high-speed chemical imaging of single VACV by reducing the acquisition time to 0.32 s per image per wavenumber of single viruses and the SNR of one single VACV is ~ 4 within the field of view (FoV) of 24 by 24 µm (Supplementary Fig. 10). In comparison, the image acquisition time of one single virus is 46.4 s at the FoV of ~2.3 by 2.3 µm in previous work [15]. Together, WIDE-MIP microscopy provides ~1000-fold higher throughput, enabling fingerprinting of viral vectors for quality control.

To use this method, the viral vector products can be prepared on a silicon substrate following the sample preparation protocol and then imaged using WIDE-MIP. The substrate will then be taken for WIDE-MIP imaging and get the fingerprints of the particles from the products. For fingerprint region from 1500 $cm^{-1}$ to 1750 $cm^{-1}$, fingerprinting at each FoV will take at least 8.32 s with the scanning step at 10 $cm^{-1}$. Quality control results can be obtained by comparing the fingerprints of the products with those from the standard viral vector, followed by spectral analysis of viral proteins and viral nuclei acids to determine stability, purity, and integrity. Additionally, WIDE-MIP can advance the development of new viral vectors by facilitating their characterization and optimization.

**Supplementary Table 1. Comparison of detection limit of MIP imaging.**

|       | Substrate | Sample      | SNR      | Throughput (beads/min) |
|-------|-----------|-------------|----------|------------------------|
| **A** [2]  | Glass     | 100 nm PS   | ~ 70     | 1                      |
| **B** [16] | CaF$_2$   | 100 nm PS   | ~ 10-50  | 1.5                    |
| **C** [15] | CaF$_2$   | 100 nm PMMA | ~ 13     | 3                      |
| **This work** | Silicon | 100 nm PMMA | ~ 21   | > 3000                 |

PS: polystyrene. PMMA: polymethyl methacrylate. SNR: signal-to-noise ratio.

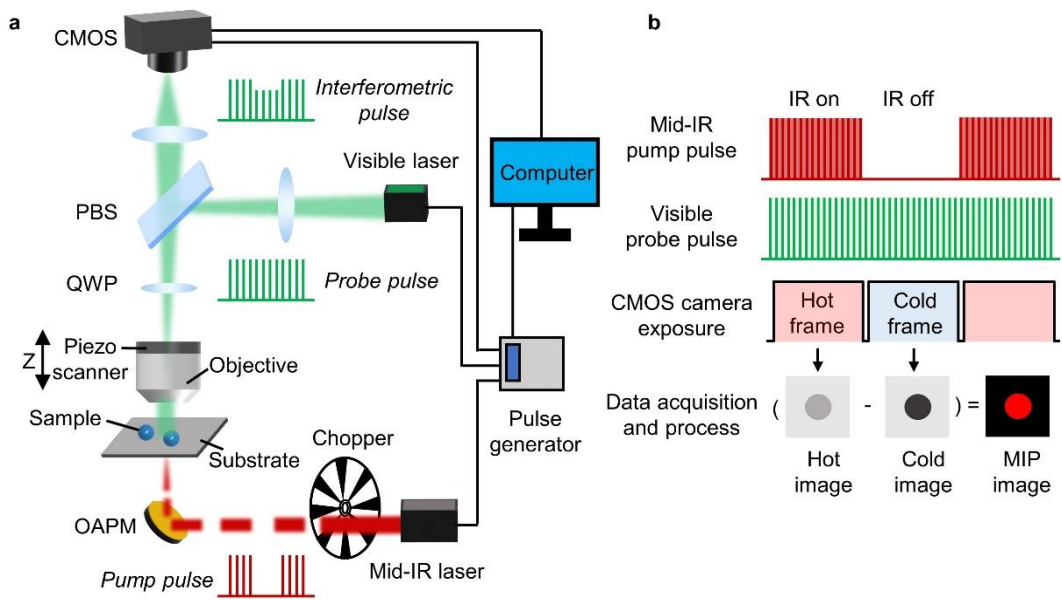

**Supplementary Fig. 1 WIDE-MIP microscope setup and signal synchronization.** (a) Schematic of WIDE-MIP microscope. The IR pump beam was generated by a tunable (from 1400 to 1800 $cm^{-1}$) mid-IR laser operating at 20 kHz repetition rate with a ~20 ns pulse duration, which was further modulated by an optical chopper. The visible probe was provided with a pulsed 520 nm nanosecond laser with a pulse duration of 129 ns. The interferometric scattering was recorded by a 2 million well-depth camera. A delay pulse generator is used to synchronize the pump pulse, probe pulse and camera. PBS: Polarizing beam splitter, OAPM: Off-axis parabolic mirror, QWP: quarter-wave plate, CMOS: complementary metal-oxide semiconductor. (b) Illustration for the synchronization and data acquisition of WIDE-MIP microscopy. To synchronize and acquire data for WIDE-MIP microscopy, the delay pulse generator was triggered by the output signal from the nanosecond IR laser. The oscilloscope was used to monitor the IR and visible pulses through a Mercury-Cadmium-Telluride detector and a photodiode, respectively. The IR pulses were modulated to a 50% duty cycle by the optical chopper. The camera trigger signal delay was adjusted to capture both IR on (hot) and IR off (cold) frames. The MIP contrast was generated by the subtraction of hot and cold frames.

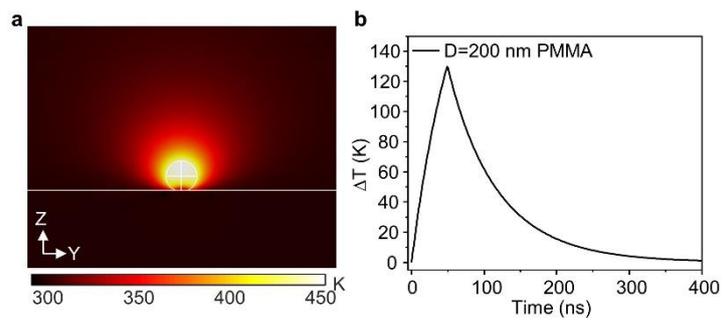

**Supplementary Fig. 2 Simulated temperature rise of a 200 nm PMMA bead under single IR pulse heating.** (a) Temperature distribution of a 200 nm PMMA bead on the silicon substrate heated by a single IR pulse. Time is at 400 ns after the rising edge of the IR pulse. (b) Simulation results of thermodynamic properties of the heated PMMA bead.

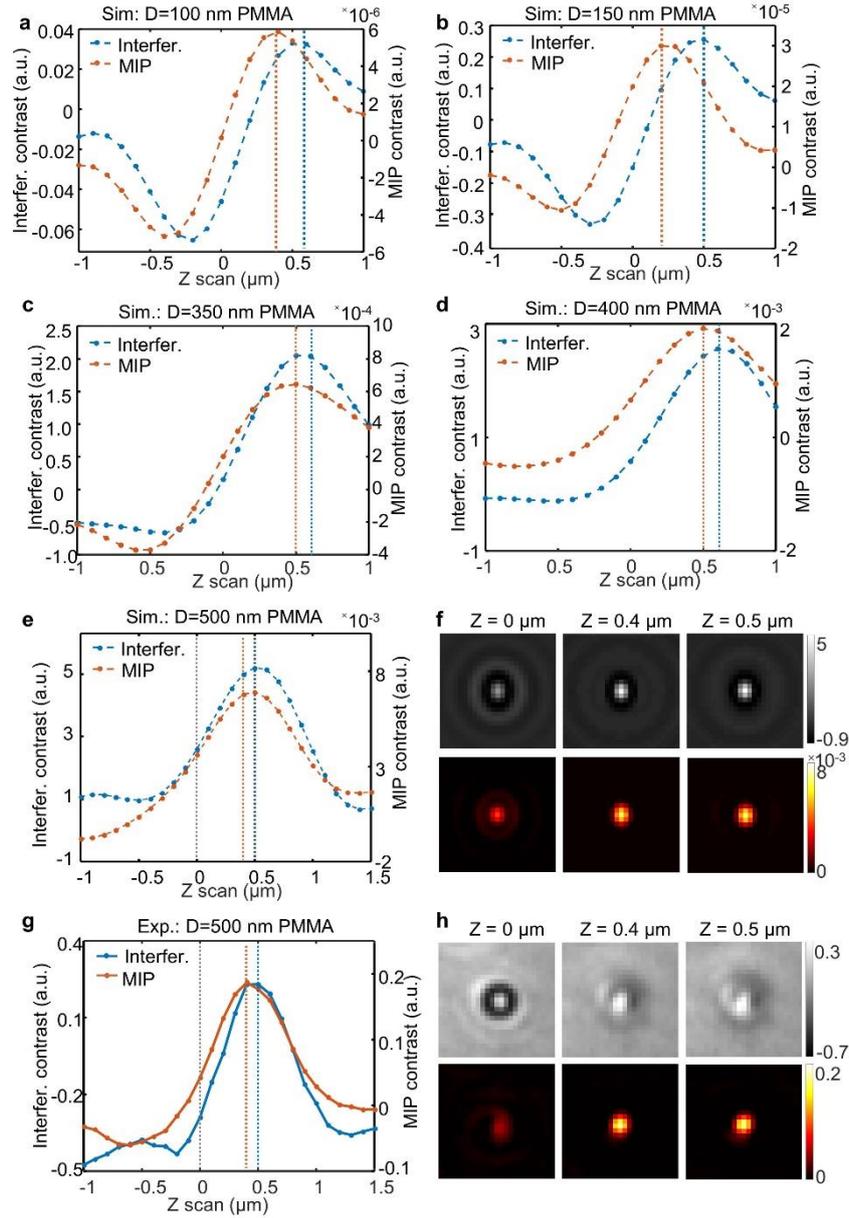

**Supplementary Fig. 3 Interferometric defocus enhancement of MIP contrasts of PMMA beads with different sizes.** Simulated defocus curves of interferometric and MIP contrasts of single (a) D = 100 nm, (b) D = 150 nm, (c) D = 350 nm, (d) D = 400 nm, (e) D = 500 nm PMMA bead. (f) Simulated interferometric and MIP images at Z = 0 μm, Z = 0.4 μm, and Z = 0.5 μm. FoV: 3 μm by 3 μm. (g) Experimental defocus curves of interferometric and MIP contrasts for D = 500 nm PMMA beads. (c) Experimental interferometric and MIP images at Z = 0 μm, Z = 0.4 μm, and Z = 0.5 μm. FoV: 3 μm by 3 μm. Red dot lines indicate the maximum MIP contrasts. Blue dot lines indicate the maximum interferometric contrasts. Grey dot lines indicate Z = 0 μm. Power before the objective: pump: 48 mW at 1728 cm$^{-1}$, probe: ~1 mW. Acquisition time: 2.36 s per image. Objective piezo scanner, Piezosystemjena, MIPOS 100, Z-axis scanning step: 100 nm. FoV: field of view. D: Diameter. Interfer.: Interferometric. $\Delta T$ was set to 1 K for a simplified simulation.

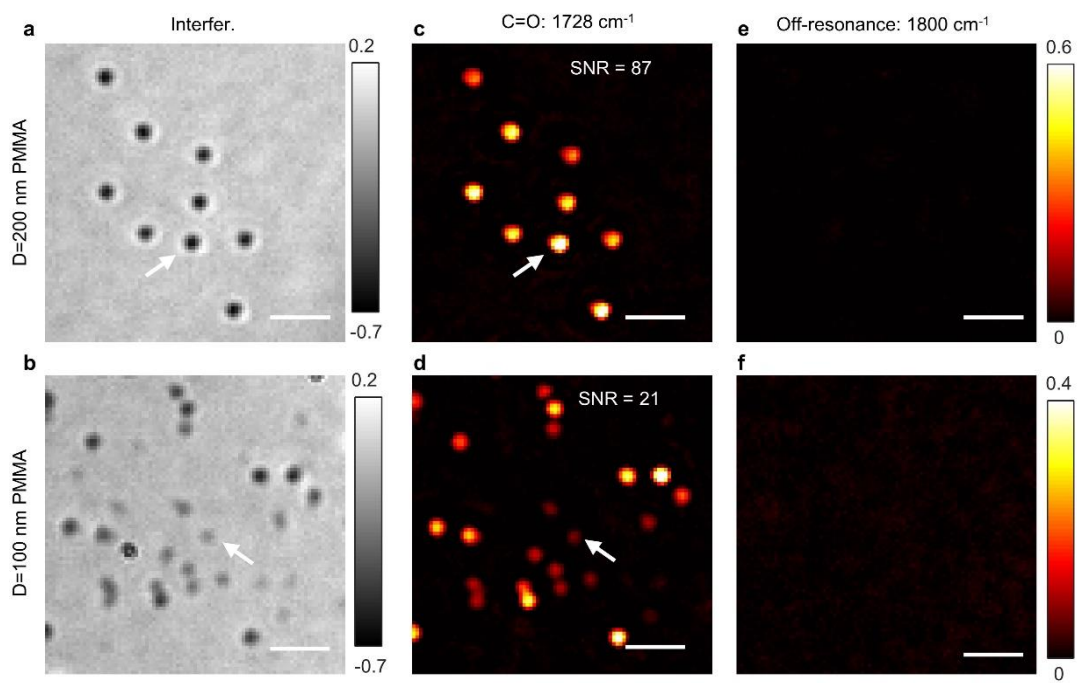

**Supplementary Fig. 4 Detection limit of WIDE-MIP imaging.** Defocused interferometric scattering image of (a) D = 200 nm and (b) D = 100 nm PMMA beads. (c, d) MIP image of the same area with the pump at 1728 cm$^{-1}$. (e, f) Off-resonance image showed no contrast. Scale bars: 2 μm. Power before the objective: pump: 31.4 mW at 1728 cm$^{-1}$, 32.2 mW at 1800 cm$^{-1}$, probe: ~1 mW. Image acquisition time: 2.36 s per image. The MIP intensities are normalized by the IR power. The larger signals in panel (d) are from particle aggregates.

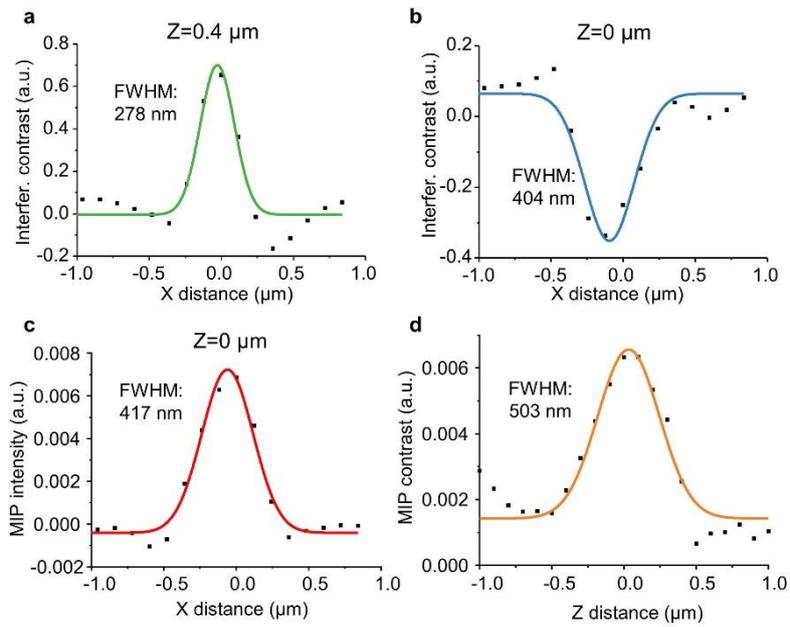

**Supplementary Fig. 5 Spatial resolution of WIDE-MIP imaging.** (a, b) Horizontal cross-sectional profiles of interferometric scattering image across one single bead (shown in Fig. 2f) at (a) Z = 0.4 μm and (b) Z = 0 μm. The Gaussian fitted FWHMs are 278 and 404 nm. (c) Horizontal cross-sectional profiles of MIP image across the same bead (shown in Fig. 2g) at Z = 0 μm. The Gaussian fitted FWHM is 417 nm. (d) Axial cross-sectional profiles of MIP image across the same bead (shown in Fig. 2g) at Z = 0 μm. The depth of focus for MIP imaging is 503 nm, calculated from the Gaussian fitted FWHM.

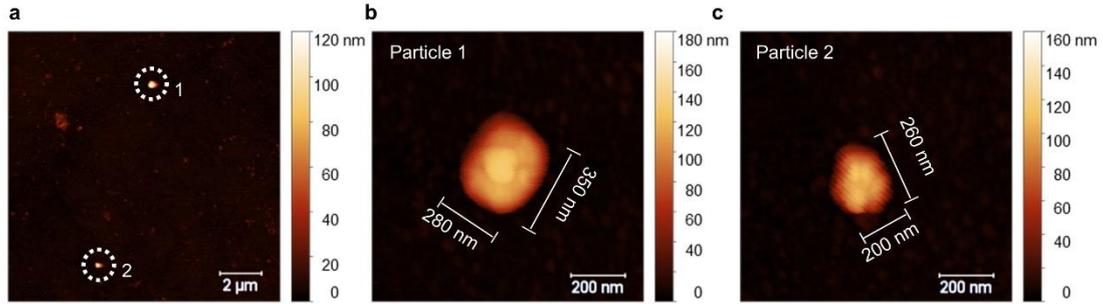

**Supplementary Fig. 6 AFM analysis of size and shape of a single VACV.** (a) AFM characterization of a single VACV virus in air. (b) and (c) are zoom-in AFM images of single viruses labeled with dotted circles in (a).

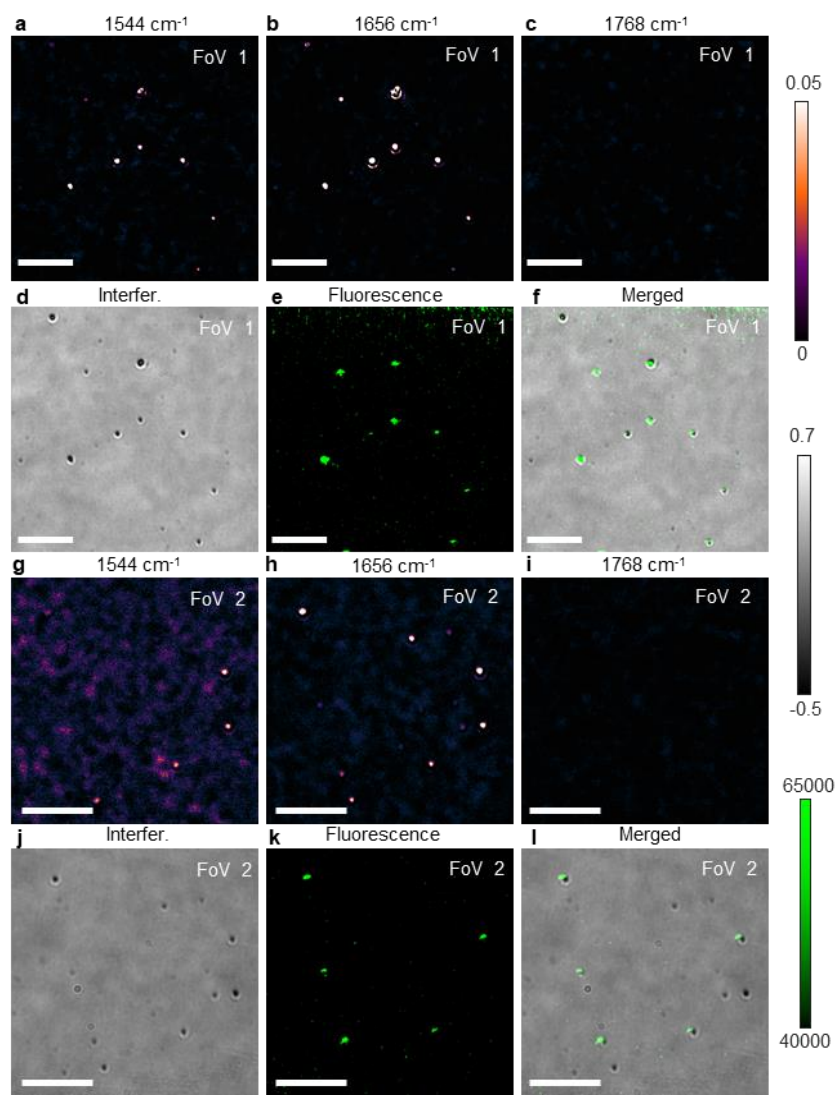

**Supplementary Fig. 7 Fingerprinting detection of single VSVs.** (a) Amide II bond-selective image of single VSV viruses with the pump at 1552 cm$^{-1}$ of FoV 1. (b) Amide I bond-selective image of the same with the pump at 1656 cm$^{-1}$. (c) Off-resonance image showed no contrast. (d) Defocused interferometric scattering, (e) fluorescence and (f) merged images of the same area. (g) Amide II bond-selective image of single VSV viruses with the pump at 1552 cm$^{-1}$ of FoV 2. (h) Amide I bond-selective image of the same with the pump at 1656 cm$^{-1}$. (i) Off-resonance image showed no contrast. (j) Defocused interferometric scattering, (k) fluorescence and (l) merged images of the same area. Scale bars: 10 μm. Power before the objective: pump: 29.1 mW at 1552 cm$^{-1}$, 34.5 mW at 1656 cm$^{-1}$, 35.8 mW at 1768 cm$^{-1}$, probe: ~1 mW. Image acquisition time: 2.36 s per wavenumber. The MIP spectrum is normalized by the IR power. FoV: field of view.

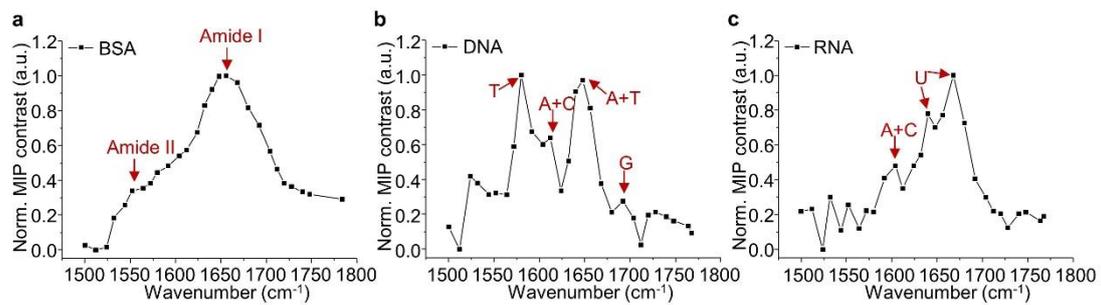

**Supplementary Fig. 8 WIDE-MIP spectra of pure chemicals.** WIDE-MIP spectra of (a) dried pure protein (BSA) film, (b) dried pure DNA film (cDNA of melanoma cell), and (c) dried pure RNA film (ssRNA of T24 cell). The MIP spectra are normalized by the IR power under each wavenumber.

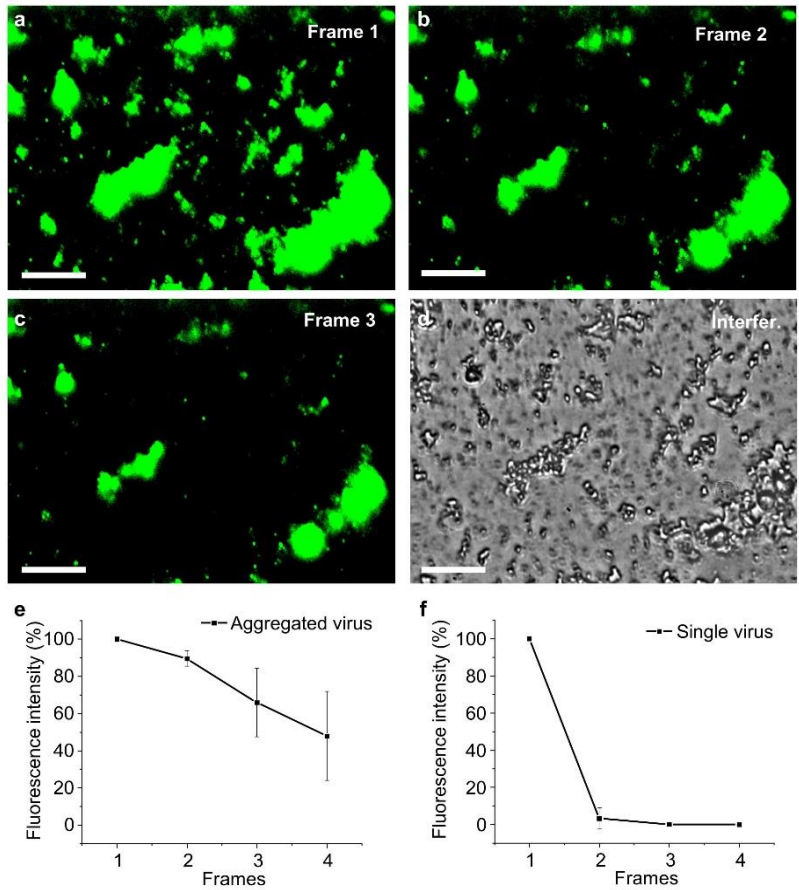

**Supplementary Fig. 9 Photobleaching analysis of single VACVs.** (a-c) Continuous fluorescence imaging of VACVs from frame 1 to frame 3. (d) Defocused interferometric scattering image of the same area in (a-c). Scale bars: 10 μm. Image acquisition time: 5 s/image. Fluorescence intensity of (e) aggregated viruses (n =15) and (f) single viruses (n = 20) in the field of view.

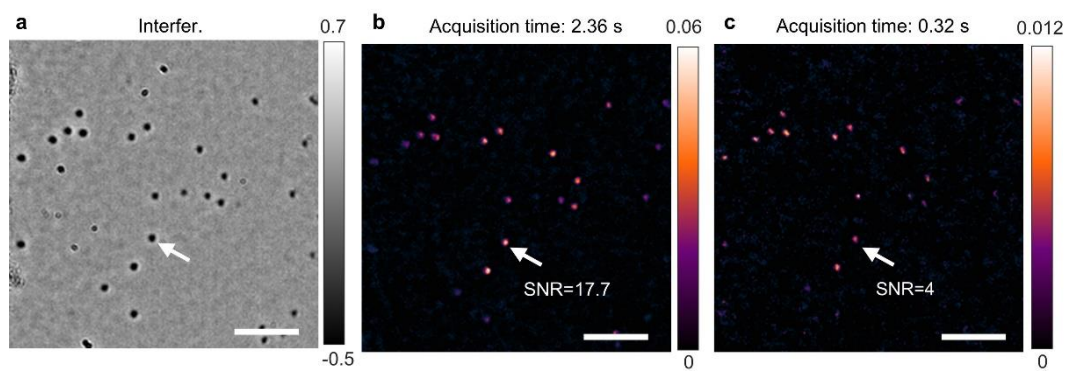

**Supplementary Fig. 10 High speed chemical imaging of single VACVs.** (a) Defocused interferometric scattering image of single VACV viruses. Amide I bond-selective image of the same area with the image acquisition times of (b) 2.36 s and (c) 0.32 s. The pump beam wavenumber is set to 1656 cm$^{-1}$. Power before the objective: 34.5 mW at 1656 cm$^{-1}$, probe: ~1 mW. Scale bars: 5 μm.

**Supplementary References**